\documentclass[aps,pre,reprint, amsmath, amssymb,superscriptaddress]{revtex4-1}

\usepackage{bm}
\newcommand{\beq}{\begin{equation}}    
\newcommand{\eeq}{\end{equation}}
\newcommand{\kB}{k_{\mbox{\tiny B}}}
\usepackage[retainorgcmds]{IEEEtrantools}
\usepackage{graphicx,tikz,placeins}
\usepackage{mathrsfs}
\usepackage{amsmath,amssymb,amsfonts,physics}
\usepackage{color}
\usepackage{times,txfonts}
\usepackage{nicefrac}
\usepackage[colorlinks=true,linkcolor=blue,urlcolor=blue,citecolor=blue,pdfusetitle]{hyperref}
\usepackage{physics}
\usepackage{soul}
\newcommand{\mom}[1]{\left\langle#1\right\rangle}

\begin{document}

\title{Obtaining efficient  thermal engines from interacting Brownian particles under time dependent periodic drivings }

\author{Iago N. Mamede}
\affiliation{Instituto de F\'{\i}sica da Universidade de S\~{a}o Paulo, 05314-970 S\~{a}o Paulo,  Brazil}
\author{Pedro E. Harunari}
\affiliation{Instituto de F\'{\i}sica da Universidade de S\~{a}o Paulo, 05314-970 S\~{a}o Paulo,  Brazil}
\affiliation{Complex Systems and Statistical Mechanics, Physics and Materials Science Research Unit, University of Luxembourg, L-1511 Luxembourg, Luxembourg}
\author{Bruno A. N. Akasaki}
\affiliation{Instituto de F\'{\i}sica da Universidade de S\~{a}o Paulo, 05314-970 S\~{a}o Paulo,  Brazil}
\author{Karel Proesmans}
\affiliation{Complex Systems and Statistical Mechanics, Physics and Materials Science Research Unit, University of Luxembourg, L-1511 Luxembourg, Luxembourg}
\affiliation{Hasselt University, B-3590 Diepenbeek, Belgium}
\author{C. E. Fiore}
\affiliation{Instituto de F\'{\i}sica da Universidade de S\~{a}o Paulo, 05314-970 S\~{a}o Paulo,  Brazil}
\date{\today} 
\begin{abstract}
We    introduce  an alternative route  for obtaining reliable cyclic engines, based on interacting Brownian particles  under time-periodic drivings. General expressions for the thermodynamic fluxes, such as power and heat, are obtained using the framework of Stochastic Thermodynamics. 
Several  protocols for optimizing the engine performance are considered, by looking at system parameters such as the output forces and their phase-difference. 
We study both work-to-work and heat-to-work engines.
 Our results suggest that carefully designed interactions between particles can lead to more efficient engines.

\end{abstract}

\maketitle

\section{Introduction}
Small scale  engines operating out of equilibrium have received a substantial increase of attention
in the last years, especially because several  process in nature (mechanical, biological, chemical and others)  are related to some kind of energy conversion (e.g. mechanical into chemical and vice-versa) \cite{seifert2012stochastic,liepelt1,liepelt2}. 
The constant fluctuating flow of energy constitutes a fundamental feature fueling the operation of nonequilibrium engines which is well described by the framework of Stochastic Thermodynamics \cite{seifert2012stochastic}.

Entropy production plays a fundamental role in Nonequilibrium Thermodynamics. It satisfies fluctuation theorems \cite{crooks,jarz} and bounds such as the Thermodynamic Uncertainty Relations (TURs) \cite{barato2015thermodynamic,pietzonka2016universal,barato2018bounds,barato2019unifying,proesmans2017discrete,harunari2020,PhysRevE.99.062126,PhysRevResearch.2.013060} and can be extended for deriving
general bounds between power, efficiency and dissipation \cite{PhysRevLett.120.190602}. Here we look at a case-study of a cyclic heat engines in which the nonequilibrium features are due to distinct thermal reservoirs and time-dependent external forces.


Brownian particles are often at the core of nano-scaled heat engines \cite{martinez2016brownian, krishnamurthy2016micrometre,blickle2012realization, proesmans2016brownian, quinto2014microscopic,jones2015optical, albay2018optical, kumar2018nanoscale, paneru2020colloidal,li2019quantifying}. Most of them  are based on single particle engines and have been studied for theoretical \cite{noa2020thermodynamics, proesmans2015efficiency, rana2014single, schmiedl2007efficiency, hoppenau2013carnot, tu2014stochastic, chvosta2010energetics, verley2014unlikely, imparato2007work, plata2020building} and experimental \cite{martinez2016brownian, albay2021shift,jun2014high} settings.
 On the other hand, the number of
 studies on the thermodynamic 
 properties of interacting chains of particles are limited and often constrained to time-independent driving \cite{park2016efficiency,akasaki2020entropy,li2019quantifying}. 
The scarcity of results, together the richness of such system,  raises distinct and relevant questions about the interaction contribution to the performance, the interplay between interaction and driving forces and choice of protocol optimization. The latter is a field in itself with a lot of recent works focusing on the optimization of distinct engines in terms of efficiency and/or power \cite{curzon1975efficiency, schmiedl2007efficiency, holubec2014exactly, PhysRevResearch.3.023194, PhysRevLett.109.190602}.

In this work we conciliate above issues by introducing an interacting version of the underdamped Brownian Duet \cite{proesmans2017underdamped},
in which each particle is subject to
a distinct thermal bath and 
driving force.
The existence of
distinct parameters (interaction between particles, strength of
forces, phase difference and frequency) provides
several routes for tackling optimization that
will be analyzed using the framework of stochastic thermodynamics. The introduction of interaction will provide additional control and also enhancement of power and/or efficiency. Distinct types
of optimization will be introduced and analyzed: maximization of output power
and efficiency with respect to the output forces, phase difference between external forces and interaction. 
 
Two different situations will be addressed. Initially, we consider the case in which the thermal baths
have the same temperature (interacting particle work-to-work converter) \cite{proesmans2016brownian}. 
We then advance beyond the work-to-work
converter by including a temperature difference between thermal
baths and general predictions are obtained for
distinct set of temperatures.

The  paper is structured as follows: in Section~\ref{model} we introduce the model and the main expressions for relevant quantities. In Section~\ref{effi}, we analyze the engine performance for distinct regime
operations. Conclusions are drawn in Section \ref{conc}.

\section{Thermodynamics of interacting Brownian engines}\label{model}

\usetikzlibrary{decorations.pathmorphing}
\begin{center}
\begin{tikzpicture}[scale=0.7]
\definecolor{cor1}{RGB}{147,149,153}
\definecolor{cor2}{RGB}{247, 225, 14}
\definecolor{cor3}{rgb}{0.87, 0.19, 0.39}
  \def\d{5}
  
  \draw[rounded corners, fill=cor1!35,draw=cor1,thick] (-2, -.5) rectangle (2, 3.5);
  \draw[thick] (-1.73,3) parabola bend (0,0) (1.73,3);
  \node at (-.5,2.8) {$V_1(x_1,x_2)$};
  \node[color=cor1!50!black] at (-1.4,0.0) {$T_1$};
  
  \draw[rounded corners, fill=cor2!35,draw=cor2,thick] (-2+\d, -.5) rectangle (2+\d, 3.5);
  \draw[thick] (-1.73+\d,3) parabola bend (0+\d,0) (1.73+\d,3);
  \node at (.5+\d,2.8) {$V_2(x_1,x_2)$};
  \node[color=cor2!50!black] at (1.4+\d,0.0) {$T_2$};
  
  \draw[->,thick] (0,1.8) -- (1,1.8);
  \node at (.5,2.2) {\(f_1(t)\)};
  \draw[->,thick] (0+\d,1.8) -- (-1+\d,1.8);
  \node at (-.5+\d,2.2) {\(f_2(t)\)};

  \draw[decorate, decoration=snake, color=cor3,thick] (0,1) -- (\d,1);
  \node[color=cor3] at (\d/2,.5) {\(\kappa\)};
  
  \shade[ball color=cor1] (0,1) circle(0.7);
  \node at (0,1) {1};
  \shade[ball color=cor2!50] (\d,1) circle(0.7);
  \node at (\d,1) {2};
\end{tikzpicture}
\end{center}

The model is composed by two interacting underdamped Brownian particles with equal mass \(m\),  each one  subject
to a distinct external force and  placed in contact with a thermal bath of temperature \(T_i\), \(i=\{ 1,2\}\). Their positions and velocities, $x_i$ and $v_i$, evolve in time according to the following set of  Langevin equations:
\begin{equation}
    \dv{v_1}{t} =  \frac{1}{m}F_1^*(x_1,x_2) +\frac{1}{m}F_1(t) -\gamma v_1 + \zeta_1,
\end{equation}
\begin{equation}
    \dv{v_2}{t} =   \frac{1}{m} F_2^*(x_1,x_2) +\frac{1}{m}F_2(t) -\gamma v_2 + \zeta_2,
\end{equation}
and
\begin{equation}
    \dv{x_1}{t}= v_1,\quad \dv{x_2}{t}=v_2,
\end{equation}
respectively.
There are eight forces acting on the system: two forces $F_i^*(x_1,x_2)$, related to the harmonic potentials and the interaction between particles,  two external driving components $F_i(t)$, 
friction forces  $-\gamma v_i$
(with \(\gamma\) denoting the friction parameter) and  stochastic forces $\zeta_i(t)$. The former can be written
as the derivative of a potential $V_i$ given by $F_i^*(x_1,x_2)=-\partial V_i / \partial x_i$, whereas the stochastic forces are described as a white noise:  \(\expval{\zeta_i(t)}=0\) and  \(\expval{\zeta_i (t) \zeta_j (t')} =  2 \gamma  k_\text{B} T_i\delta_{ij} \delta (t-t')/m  \).
The above set of Langevin equations are associated with the probability distribution \(P(x_1,x_2,v_1,v_2,t)\) having its time evolution governed by Fokker-Planck-Kramers (FPK) equation:
\begin{equation}
\frac{\partial P}{\partial t} = - \sum_{i=1}^2\left(
v_i\frac{\partial P}{\partial x_i}
+ [F_i^*+F_i(t)] \frac{\partial P}{\partial v_i}
+ \frac{\partial J_i}{\partial v_i}\right),
\label{FPK}
\end{equation}
where
\begin{equation}
J_i = - \gamma v_i P - \frac{\gamma \kB T_i}{m}
\frac{\partial P}{\partial v_i}.
\label{3a}
\end{equation}


If the temperatures of both particles  are equal
and the external forces are absent, the probability distribution
approaches  for large times the Gibbs equilibrium distribution,
$P^\text{eq}(x_1,x_2,v_1,v_2) \propto e^{-E/\kB T}$, 
where $E=\sum_i (mv_i^2/2 + V_i)$ is the total energy of the system.
From now on, we shall consider
harmonic potentials $V_i=k_ix_i^2/2+\kappa(x_i-x_j)^2/2$,
whose associate forces
read $F_i^*=-kx_i-\kappa(x_i-x_j)$. 
The time evolution  of a generic average \(\expval{x_i^n v_j^m}\) can be obtained from the FPK equation, Eq. (\ref{FPK}), and performing appropriate partial integrations by  assuming that \(P(x_1,x_2,v_1,v_2,t)\) and its derivatives vanish when \(x_i\) or \(v_i\) approaches to \(\pm\infty\).
More specifically, we are interested in
obtaining expressions for thermodynamic quantities, such as the heat exchanged between particle $i$ and 
the reservoir and  the work rate performed by each external force over its particle. Their expressions can be obtained from the time evolution of mean energy $\langle E \rangle$ together the FPK equation and assumes a form consistent with  the first law of Thermodynamics \cite{tome2010entropy,seifert2012stochastic,broeck2010a}:
\begin{equation}
\frac{d\langle E \rangle}{dt} = - \sum_{i=1}^{2}(\dot{W}_i+\dot{Q}_i),
\label{10}
\end{equation}
where  $\dot{W}_i$ is work done over particle $i$, due to the external force $F_i(t)$,
\begin{equation}\label{work}
    \dot{W}_i = -mF_i(t) \expval{v_i},
\end{equation}
and $\dot{Q}_i$ is the heat delivered to reservoir $i$. An expression for the heat can be derived from the above two equations:
\begin{equation}\label{heat}
    \dot{Q}_i = \gamma\left( m\expval{v_i^2}-k_\text{B} T_i \right).
\end{equation}
Similarly, the time evolution of system entropy $S=-\kB\langle \ln P(x_1,x_2,v_1,v_2)\rangle$ is the difference between
entropy production  rate \(\sigma\) and entropy flux rate \(\Phi\) to/from the system to/from the thermal reservoir given by \cite{tome2010entropy,seifert2012stochastic,broeck2010a}
\beq
\sigma = \frac{m}{\gamma } \sum_{i=1}^2 \frac{1}{T_i}\int \frac{J_i^2}{P}dx_1dx_2dv_1dv_2,
\eeq
and
\beq
\Phi = -\sum_{i=1}^2 \frac{m}{T_i}\int v_i J_i dx_1dx_2dv_1dv_2,
\label{140}
\eeq
respectively. Note that $\sigma\ge 0$ (as expected), whereas $\Phi$ can be conveniently rewritten in terms of the ratio between $\dot{Q}_i$ and the temperature $T_i$:
\beq
\Phi = \sum_i \gamma \left( \frac{m\langle v_i^2\rangle}{T_i} - k_\text{B}\right)= \sum_{i=1}^2 \frac{\dot{Q}_i}{T_i}.
\label{14}
\eeq

 It is  convenient to  relate  averages $\langle v_i\rangle$'s and $\langle v_i^2\rangle$'s by means their covariances
$b_{ij}^{vv}(t)\equiv \langle v_iv_j\rangle(t)-\langle v_i\rangle(t)\langle v_j\rangle (t)$. For simplifying matters, from now on we set $m=k_B = 1$. Due
to the interaction between particles,  $b_{ij}^{vv}(t)$ also
depends on covariances $b_{ij}^{xx}(t)$'s and  $b_{ij}^{xv}(t)$'s ($x$ and $v$ attempting to the position and velocity of the $i$-th and $j-$th particles, respectively). 
Their time evolutions are straightforwardly obtained
from Eq. (\ref{FPK}), whose expression for $b_{11}^{vv}$ is given by 
\begin{equation}
   b_{11}^{vv} = \frac{T_{1} + T_{2} }{2 } +\frac{\left( T_{1} - T_{2} \right)}{2} \frac{\gamma^2 (\kappa+k) }{ \left[\kappa^{2} +  \gamma^{2}(\kappa+ k) \right]},
\end{equation}
 and  $b_{22}^{vv}$ is obtained just by exchanging $1\leftrightarrow 2$. 

\subsection{Periodically driving forces}\label{ext}
Having obtained the general expressions for a chain
of two interacting particles, we are now
in position to get  expressions in the presence of external forces. Our aim is to study the effect that interactions have on the performance of an engine. To do this, we will focus  on the simplest case in which particles are subject to  harmonic time-dependent forces $F_i(t)$ of different amplitude,  same frequency $\omega$, but  with a lag \(\delta\) between them \cite{proesmans2016brownian,proesmans2017underdamped,fiore2019,akasaki2020entropy}
\begin{equation}\label{ext1}
   F_1(t)= X_1 \cos{(\omega t )},
\end{equation}
and
\begin{equation}\label{ext2}
  F_2(t,\delta)= X_2 \cos{[\omega (t-\delta )]},
\end{equation}
respectively. 
The system will relax to a time-periodic steady state with 
 $\overline{\dot{Q}}_1+\overline{\dot{Q}}_2=-(\overline{\dot{W}}_1+\overline{\dot{W}}_2)$, where
 each  mean work $\overline{\dot{W}}_i$ and 
heat ${\overline {\dot Q}_i}$  are given by \begin{equation}\overline{\dot{W}}_i=-\frac{\omega}{2\pi}\int_{0}^{2\pi/\omega} F_i(t)\langle v_i\rangle(t) dt,
\label{work1}
\end{equation} and 
\begin{eqnarray}
  {\overline {\dot Q}_i}&=&\frac{\omega\gamma}{2\pi}\int_{0}^{2\pi/\omega} \langle v_i\rangle^2 dt-\overline{\kappa}(T_i-T_j),\label{hea}
\end{eqnarray}
respectively,  where  $\overline{\kappa}$ is the thermal conduction given by
$\overline{\kappa}=\gamma \kappa^2/[2\kappa^2+2\gamma^2(\kappa+k)]$ \cite{welles,tome2010entropy}. 
The steady entropy production over a cycle is promptly
obtained from Eq.~(\ref{14})  and it is 
related with average work and heat according to the expression:
\begin{equation}
   \overline{\sigma}=\frac{4T^2}{4T^2-\Delta T^2}\left[ -\frac{1}{T}(\overline{\dot{W}}_1 + \overline{\dot{W}}_2) +  (\overline{\dot{Q}}_1 - \overline{\dot{Q}}_2) \frac{\Delta T}{2T^2}\right],
   \label{eq:phi}
\end{equation}
where $T=(T_1+T_2)/2$ and 
$\Delta T=T_2-T_1$. It can also be viewed as sum of two components:  ${\overline \sigma}= \Phi_T+{\overline \Phi_f}$, where the former, $\Phi_T$, due to the difference of temperatures is given by
\begin{equation}
   \Phi_T=\frac{4\overline{\kappa}\Delta T^2}{4T^2-\Delta T^2},
    \label{phit}
\end{equation}
 and the latter, due to the external forces, is given by
\begin{equation}
\label{epf}
{\overline \Phi_f}={\tilde L}_{11}X_1^2+({\tilde L}_{12}+{\tilde L}_{21})X_1X_2+{\tilde L}_{22}X_2^2, 
\end{equation}
respectively. 
Above expressions are exact and hold beyond linear regime (large forces
and/or large difference of temperatures) between thermal baths.
In order to relate them with thermodynamic fluxes and forces, 
we are going to perform the analysis of a small temperature
difference 
$\Delta T$ between thermal baths. In such case, we introduce
the forces $f_1=X_1/T$, $f_2=X_2/T$ and $f_T=\Delta T/T^2$, in such a way that
\begin{equation}
   \overline{\sigma}\approx J_1f_1+J_2f_2+J_Tf_T,
   \label{eq:phi2}
\end{equation}
where  flux $i$ ($i=1,2$ or $T$) is associate with force $f_i$ and given by the following
expressions $\overline{\dot{W}}_1=-TJ_1f_1$, $\overline{\dot{W}}_2=-TJ_2f_2$ and $\overline{\dot{Q}}_1 - \overline{\dot{Q}}_2=2J_Tf_T$. From them,
one can obtain Onsager coefficients $J_1=L_{11}f_1+L_{12}f_2$, $J_2=L_{21}f_1+L_{22}f_2$ and $J_T=L_{TT}f_T$, whose main expressions
are listed below
\begin{widetext}
\begin{equation}
    L_{11}  =L_{22}=\left(\frac{T\gamma   \omega ^2}{2}\right)\frac{  \gamma^2 \omega^2 + \left(\omega^2 -(k+ \kappa)  \right)^2 + \kappa^2 }{\left[\gamma^2\omega^2 +\left(\omega^2 - k\right)^2  \right]  \left[ \gamma^2\omega^2 + (\omega^2 - (k+2\kappa))^2 \right]},
    \label{l11}
\end{equation}
\begin{equation}
    L_{12}=\left(\frac{T\kappa   \omega}{2}\right)\frac{2 \gamma  \omega  \left(\kappa +k- \omega ^2\right)\cos (\delta  \omega )-  \left[\gamma^2\omega^2 - (\omega^2 - (k+\kappa))^2 + \kappa^2  \right] \sin (\delta  \omega )}{\left[\gamma^2\omega^2 +\left(\omega^2 - k\right)^2  \right]  \left[ \gamma^2\omega^2 + (\omega^2 - (k+2\kappa))^2 \right]},
       \label{l12}
\end{equation}
\begin{equation}
    L_{21}=\left(\frac{T\kappa   \omega}{2}\right)\frac{2 \gamma  \omega  \left(\kappa +k- \omega ^2\right)\cos (\delta  \omega )+   \left[\gamma^2\omega^2 - (\omega^2 - (k+\kappa))^2 + \kappa^2  \right]\sin (\delta  \omega )}{\left[\gamma^2\omega^2 +\left(\omega^2 - k\right)^2  \right]  \left[ \gamma^2\omega^2 + (\omega^2 - (k+2\kappa))^2 \right]},
      \label{l21}
\end{equation}
and
\begin{equation}
    L_{TT}={\overline \kappa}T^2,
\end{equation}
\end{widetext}
respectively. All other Onsager coefficients are zero.
We pause to make some comments: First, for $\Delta T=0$,  expressions for $L_{ij}$'s ($i=1$ and 2) are exact and valid for arbitrary large values of $f_i$'s. Second,  
one can verify that  $L_{11} = L_{22}\geq 0$ and \((L_{12}+L_{21})^2 \leq 4 L_{11} L_{22}\) in agreement with the second law of thermodynamics. Above conditions are promptly verified 
for all $k,\kappa$ and $\omega$. The non-diagonal Onsager coefficients \(L_{12}\) and \(L_{21}\) are not the same, except for the lagless case $\delta=0$.  Third, in the regime of low and large  frequencies, all coefficients behave as $\omega^2$ and $1/\omega^2$ (diagonal) and $1/\omega^4$ (non-diagonal for $\delta=0$), respectively. Fourth,  the non-diagonal  coefficients vanish for sufficiently weak interactions while the diagonal is finite, consistent  with  a quasi-decoupling between particles. Conversely, when the coupling parameter is very strong, \(\kappa\to\infty\), all coefficients remain finite and coincide with those for one Brownian particle in a harmonic potential subjected to both external forces.
Fifth, for large $\Delta T$,  Eq. (\ref{hea}) states that the heat exchanged with the thermal bath \(i\) has two contributions:  the first, coming from external forces, has the form $A_if_i^2+B_if_if_j+C_if_j^2$ (with coefficients
 $A_i,B_i$ and $C_i$ listed in Appendix \ref{appb}) and it is strictly non-negative. Hence, coefficients satisfy $A_i\ge 0$ and $C_i\ge 0$ and  $B_i^2-4A_iC_i\le 0$. The second term, coming from the difference
of temperatures, can be positive or negative depending on the sign of $T_j-T_i$. In the absence of external forces, the entropy production reduces to Eq. (\ref{phit}). 
Sixth,   expressions for coefficients ${\tilde L}_{ij}$'s  appearing in Eq. (\ref{epf}) (see Appendix \ref{appb}) are exact and hold beyond linear regime listed (large forces
and/or large difference of temperatures) between thermal baths.in Appendix \ref{appb}.
Seventh and last, the interplay between both terms can change the direction of the heat flowing per cycle, implying that the coupling parameter can  change the regime of operation of the engine, from heater to heat engine and vice-versa, as $\kappa$ is increased
and decreased. 
Similar findings have also been observed for two coupled double-quantum-dots \cite{PhysRevE.104.014149} and coupled spins \cite{huang2014quantum}.

\begin{figure*}[ht]
    \centering
    \includegraphics[width=.9\textwidth]{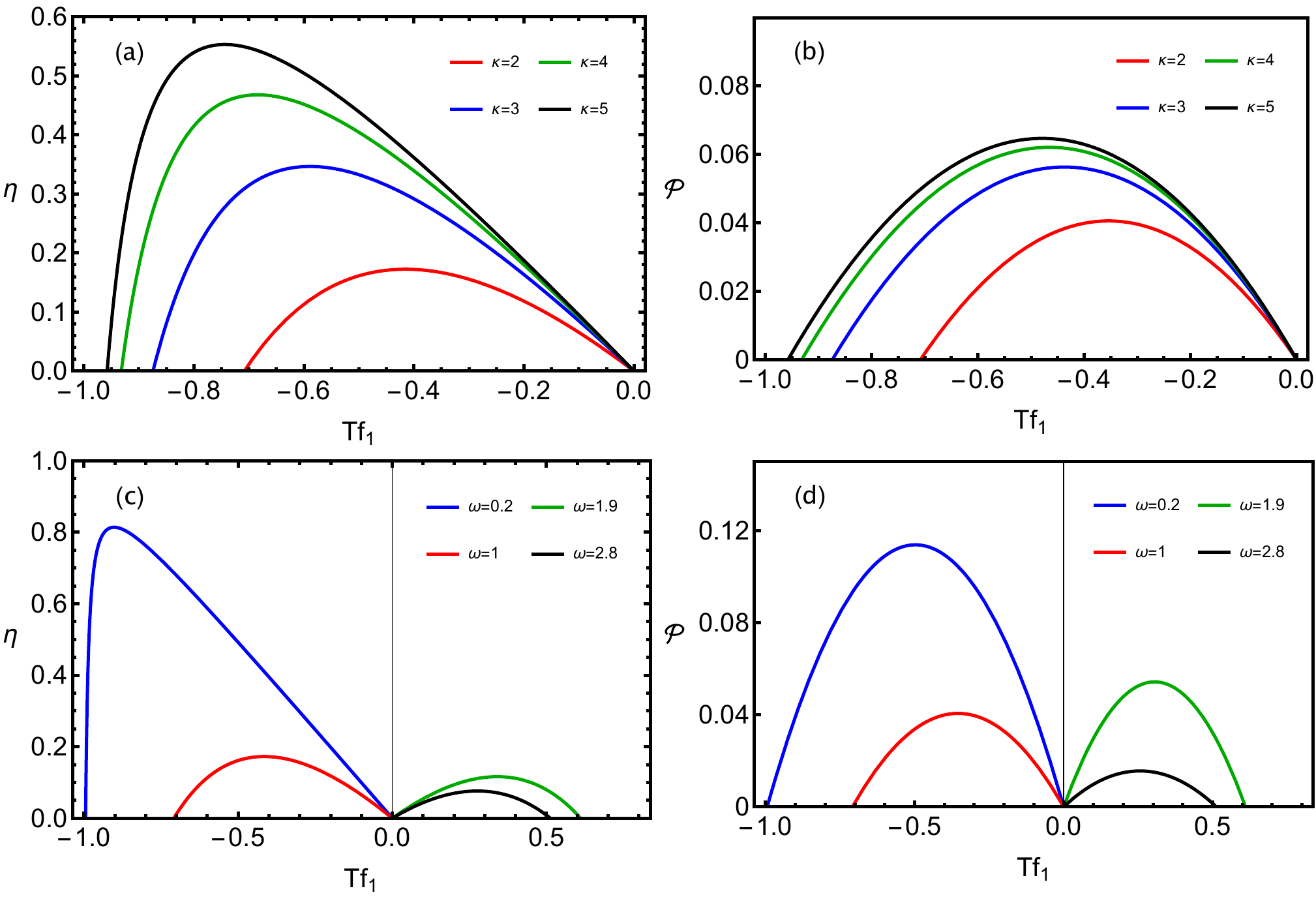}
    \caption{Panels $(a)$ and $(b)$ panels depict the
    efficiency $\eta$ and power output ${\cal P}$ versus $Tf_1$ for
    distinct $\kappa$'s and $\omega=1$. In $(b)$ and $(d)$, the same
    but for distinct $\omega$'s and $\kappa=2$. In all
    cases, we set $Tf_2=1,T=0.3,\delta=0$ and $k=0.1$.}
    \label{fig1}
\end{figure*}

\begin{figure}[ht]
    \centering
    \includegraphics[width=.4\textwidth]{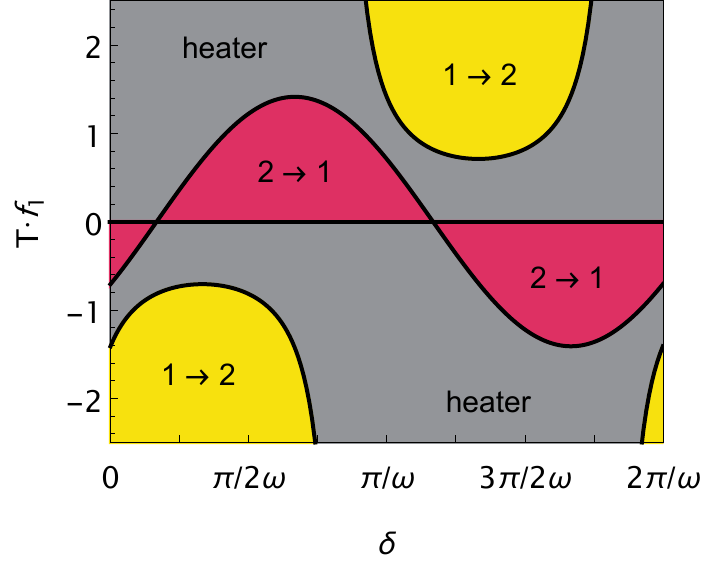}
    \caption{Phase diagram $Tf_1$ versus  $\delta$ for the work-to-work converter.  $1\rightarrow 2$/ $2\rightarrow 1$  and heater correspond to the (engine) regime in which there is the conversion 
    from $\overline{\dot{W}}_1<0$ into $\overline{\dot{W}}_2>0$/vice-versa and $\overline{\dot{W}}_1>0$ and $\overline{\dot{W}}_2>0$, respectively.  Parameters: \(Tf_2= \gamma =\omega =1\), \(k=0.1\), \(T=0.3\) and \(\kappa=2\).}
    \label{fig2}
\end{figure}

\begin{figure*}[ht]
  \centering
  \includegraphics[width=\textwidth]{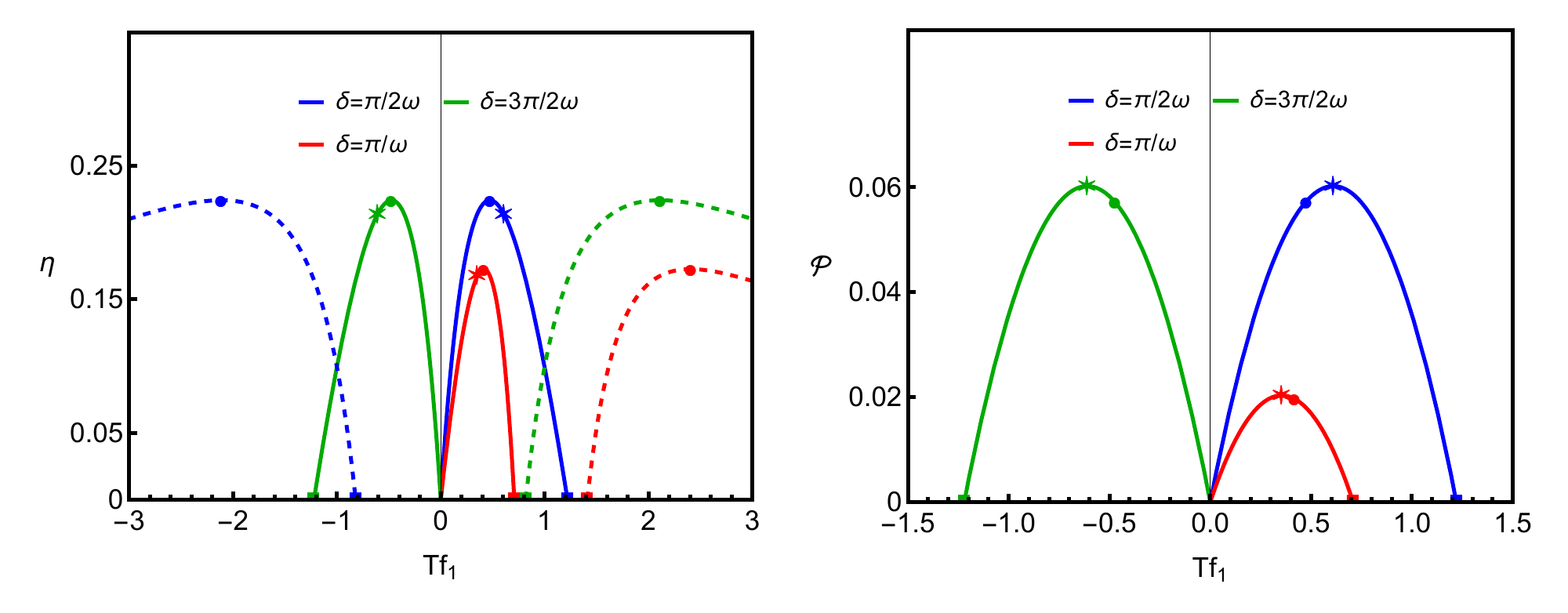}
\caption{For the same parameters from
  Fig. \ref{fig1}, the efficiency $\eta$ (left)
  and power output ${\cal{P}}$ (right) versus $Tf_1$ for distinct phase differences
  $\delta$'s. Dashed and continuous lines in left panel correspond to the conversion from $\overline{\dot{W}}_1$ into  $\overline{\dot{W}}_2$ and vice-versa, respectively. Circles, stars and squares denote the maximum efficiency, maximum power and $Tf_m$, respectively.}
    \label{fig3}
\end{figure*}

\section{Efficiency}\label{effi}
A generic  system operates as an engine when parameters  are set in such a way that
   a given amount of  energy received  is partially converted into  power output ${\cal P}\ge 0$.   A measure for  the efficiency $\eta$  is given by  the ratio between above quantities and constitutes a   fundamental quantity for characterizing  such conversion. 
Our aim here consists of exploring the role of distinct parameters, mainly the interaction between particles, 
in such a way that such system can operate as an efficient engine.   By considering for instance the particle $i=2$ as the worksource, the engine regime  implies that ${\cal P}={\overline {\dot W_{\rm 1}}}\ge 0$ and according to
 Eq. (\ref{hea})  the system will receive heat   when
 $T_1>>T_2$ ($T_2>>T_1$), consistent with ${\overline {\dot Q_{\rm 1}}}<0$ (${\overline {\dot Q_{\rm 2}}}<0$). Conversely,
when the difference of temperatures between thermal baths is small and/or
 when 
 forces $f_1/f_2$ are large,  both particles do
 not necessarily receive  heat  from the thermal bath and  only input work (actually input power) can be converted into
 output work. Such class of engines, also
 known as work-to-work converter, will be analyzed next.

We shall split the analysis in the regime of equal and different
temperatures. For both cases, we will investigate the  machine performance
with respect to the loading force $f_1$ and other parameters, such as
interaction $\kappa$ and phase difference $\delta$.

\subsection{work-to-work converter}\label{wtw}
Since for equal temperatures $\overline{\dot{Q}}_{1}$ and $\overline{\dot{Q}}_{2}$ are non negative, consistent with the system solely delivering heat to the thermal baths, Eq. (\ref{eff1})  reduces
to the ratio between worksources:
\begin{equation}
    \eta \equiv - \frac{{\cal P}}{\overline{\dot{W}}_\text{2}}=-\frac{L_{11}f_1^2+L_{12}f_1f_2}{L_{21}f_2f_1+L_{22}f_2^2},
    \label{efff}
\end{equation}
where  the second right side of Eq. (\ref{efff}) was re-expressed in terms of Onsager coefficients and thermodynamic forces.

 Fig. \ref{fig1} depicts, for $\delta=0$, the main features of the efficiency and power
output by analyzing the influence of interaction
$\kappa$ and frequency $\omega$. 
We find that the interaction between particles
improves substantially the machine performance. Properly tuning $\kappa$   not only changes 
the operation regime, from heater to a work-to-work converter (engine), but also
 increases the power, efficiency and the range of operation 
[e.g. the possible values of \(f_1\) within the same engine regime, cf. panels (a) and (b)]. Unlike the engine, in the  heater operation mode (often called dud engine), work is extracted from both worksources (\(\overline{\dot{W}}_1\) and \(\overline{\dot{W}}_2>0\)).  Contrariwise, the increase of frequency (lowering the driving period) reduces the machine efficiency. This can be  understood by the fact that the system presents some inertia and does not properly respond to abrupt changes when frequency is large.
The output force $f_1$ has opposite direction to  $f_2$ when $k+\kappa>\omega^2$ and vice-versa, as depicted in panels (c) and (d).
 
\begin{figure*}[ht]
\centering
  \includegraphics[width=.9\textwidth]{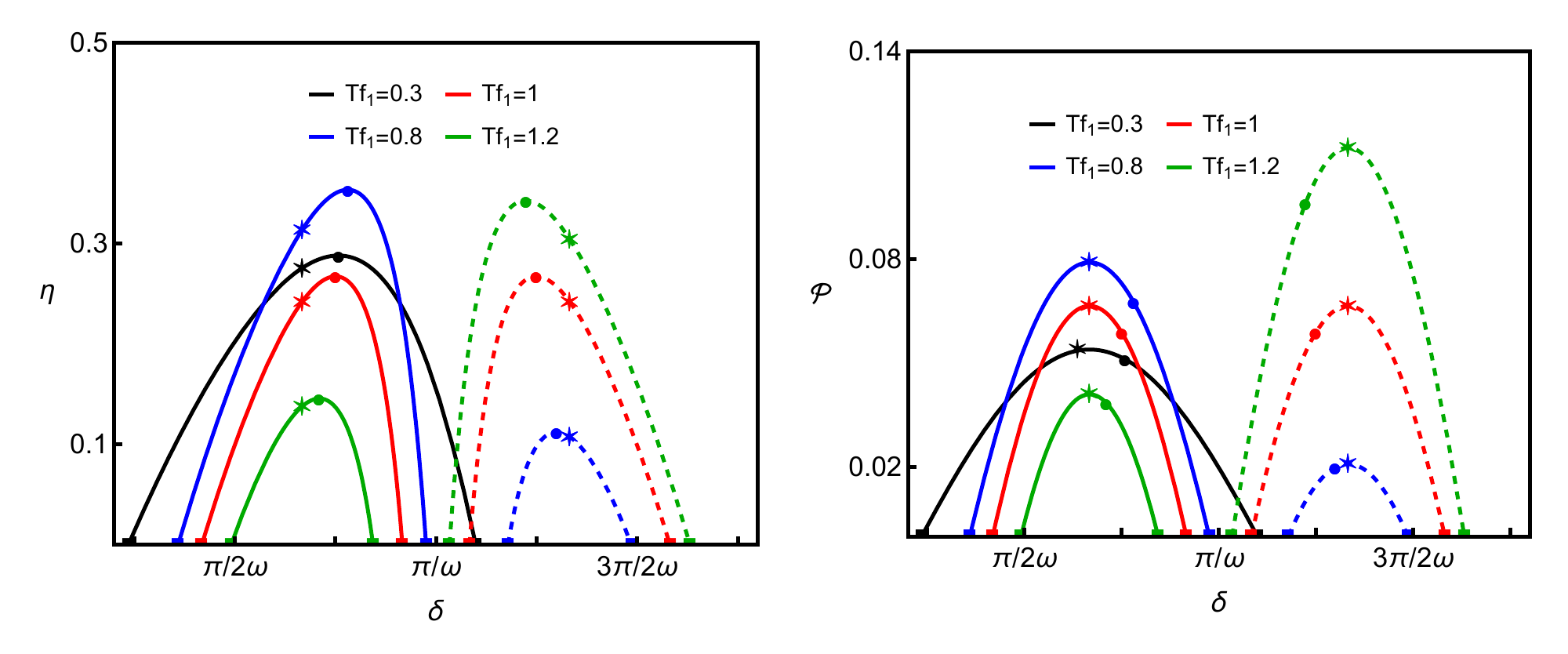}
\caption{For the same parameters from
  Fig. \ref{fig2}, the efficiency $\eta$ (left)
  and power output ${\cal P}$ (right) versus phase difference $\delta$ for distinct $Tf_1's$. Continuous and dashed lines correspond to the conversion from $\overline{\dot{W}}_2$ into  $\overline{\dot{W}}_1$ and vice-versa, respectively. Squares,  stars and circles denote the $\delta_{m1}/\delta_{m2}$, maximum power and maximum efficiency, respectively.}
  \label{fig45}
\end{figure*}

 Next, we examine the influence of a phase
 difference between harmonic forces,
 as depicted in Figs \ref{fig2}-\ref{fig45}. 
The existence of a lag between driving forces not only   controls the power and efficiency, but can also
 guide the operation modes of the system.
 In other words, depending on the
 value of $\delta$, the work is extracted from the  worksource 1 and dumped into the  worksource 2 ($\eta=-\overline{\dot{W}}_2/\overline{\dot{W}}_1$) or vice-versa ($\eta=-\overline{\dot{W}}_1/\overline{\dot{W}}_2$), 
 both conversions are possible for the same output force or even none of them.
   Such changes of conversion in the operation
 model (see e.g. Fig.~\ref{fig2}) share some similarities with some theoretical models for kinesin in which the range chemical
 potentials and mechanical forces can rule the energy
 conversion (chemical into mechanical and vice-versa) \cite{liepelt2}.  
 
 Once introduced the main features about the model parameters and how they influence the machine performance, we are going to present distinct protocols for optimizing them.

 \subsubsection{Maximization with respect to the output force}\label{maxf}
 The first (and simplest) maximization is carried out 
 with respect to the output force $f_1$  and  the other  parameters are held fixed.
Such optimizations have been performed in  Refs.~\cite{karel2016prl,noa2020thermodynamics}. 
Since ${\cal P}=\overline{\dot{W}}_\text{1}\ge 0$ the engine regime is delimited by the interval $0\le \abs{f_{1}}\le \abs{f_m}$ where $f_{m}\equiv -L_{12}f_2/L_{11}$. By adjusting
the output forces  $f_{1mP}$ and $f_{1mE}$ ensuring maximum power ${\cal P}_{mP}$  (with efficiency ${\eta}_{mP}$)
and maximum efficiency ${\eta}_{mE}$   (with power ${\cal P}_{mE}$), we obtain the following expressions, expressed in terms
of Onsager coefficients \cite{karel2016prl}:
\begin{equation}
  f_{1mE}=\frac{L_{22}}{L_{12}}\left(-1+ \sqrt{1-\frac{L_{21} L_{12}}{L_{22}L_{11}} }\right)f_2,
  \label{eq:x2meta}
\end{equation}
and
\begin{equation}
  f_{1mP}=-\frac{1}{2}\frac{L_{12}}{L_{11}}f_2,
  \label{x2mp}
\end{equation}
respectively, with corresponding efficiencies
\begin{equation}
  \eta_{mE,f_1}=-\frac{L_{12}}{L_{21}}+\frac{2L_{11}^2}{L_{21}^2}\left(1-\sqrt{1-\frac{L_{21} L_{12}}{L_{11}^2 }}\right),
  \label{etame}
\end{equation}
and
\begin{equation}
  \eta_{mP,f_1}=\frac{L_{12}^2}{4 L_{11}^2-2L_{21}L_{12}},
\label{etamp}
\end{equation}
respectively, where the property $L_{22}=L_{11}$ has been used.  Similar expressions are obtained for  ${\cal P}_{mE}$
and  ${\cal P}_{mP}$ by inserting $f_{1mE}$ and  $f_{1mP}$ into the relation
for ${\cal P}$. Maximum efficiencies are not independent from each other, but related
via simple relation
\begin{equation}
  \eta_{mP,f_1}=\frac{P_{mP,f_1}}{2P_{mP,f_1}-P_{mE,f_1}}\eta_{mE,f_1}.
\end{equation}
respectively \cite{karel2016prl}. Expressions for maximum quantities are depicted in Fig. \ref{fig3} and Fig. \ref{fig5} (continuous lines).
 


\subsubsection{Maximization with respect to the interaction or phase difference between harmonic forces}
\begin{figure}[ht]
    \centering
    \includegraphics[width=.45\textwidth]{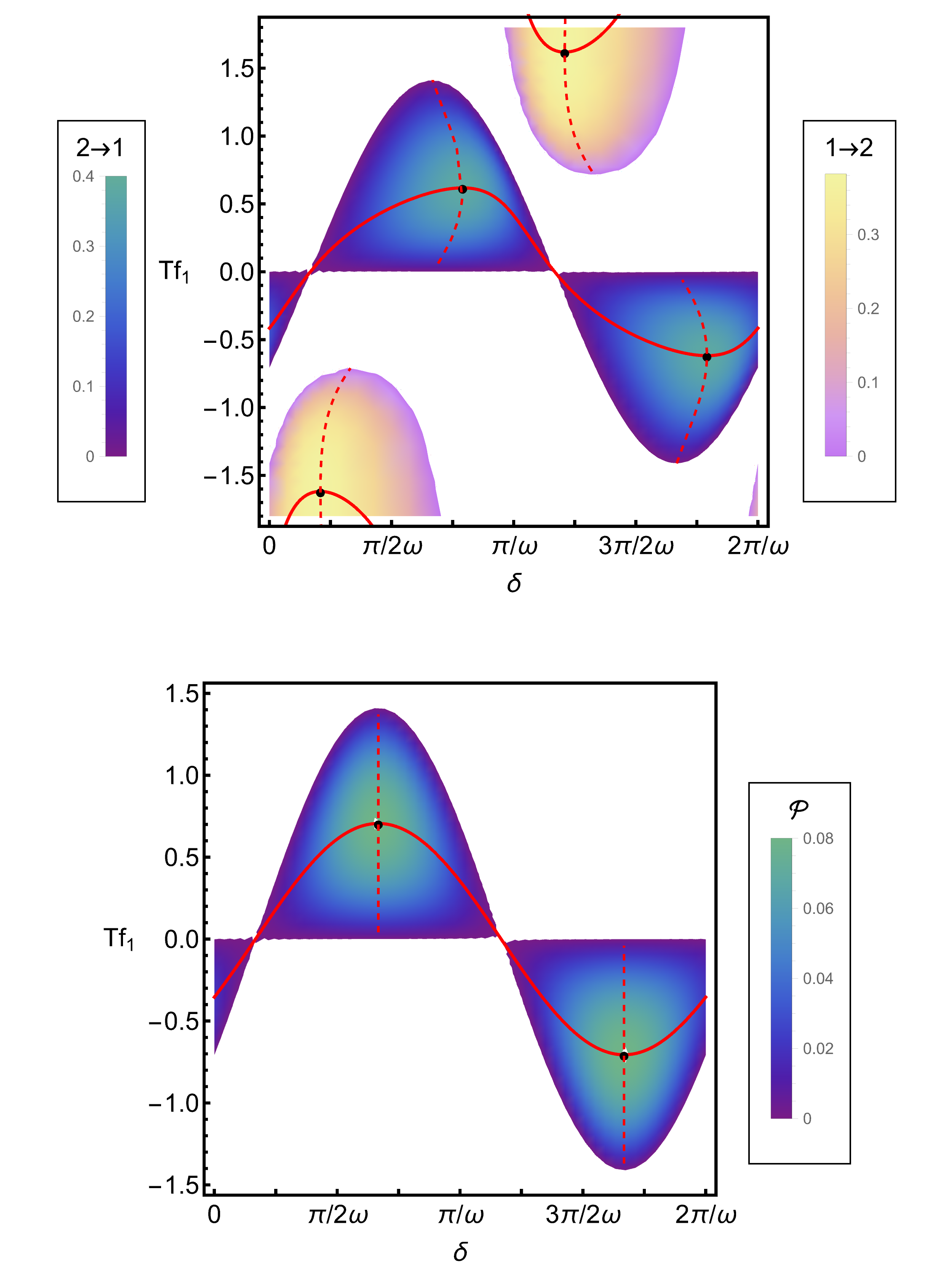}
    \caption{ For the same parameters from Fig. \ref{fig2},  depiction of  efficiency (top) and power output (bottom) for distinct $Tf_1$'s and $\delta$'s.  Continuous and dashed lines denote
    the maximization with respect to the force $f_1$ and $\delta$, respectively. The intersection between curves corresponds to the simultaneous maximization (circle).
    }
    \label{fig5}
\end{figure}

Here we present an alternative route for improving  the engine performance, based on optimal choices of $\kappa$ or $\delta$. 
Since both of them appear only in
 Onsager coefficients, their  maximizations are described by common set of relations, when expressed
 in terms of Onsager coefficients. Let $\alpha_{mP}$ and $\alpha_{mE}$ the optimal parameter ($\kappa$ or $\delta$) which maximize the power output and efficiency, respectively. From  expressions for $\cal{P}$ and $\eta$, their values are given by
\begin{equation}
f_{1}=-\frac{L_{12}^{'}(\alpha_{mP})}{L_{11}^{'}(\alpha_{mP})}f_2,
\label{lagpow}
\end{equation}
and
\begin{equation}
  f_1=\left(\frac{-B(\alpha_{mE})\pm\sqrt{B^2(\alpha_{mE})-4A(\alpha_{mE})C(\alpha_{mE})}}{2A(\alpha_{mE})}\right)f_2,
  \label{kame}
\end{equation}
 respectively, where parameters $A,B$ and $C$ are given by
\begin{equation}
  A(\alpha_{mE})=L_{11}'(\alpha_{mE})L_{21}(\alpha_{mE})-L_{11}(\alpha_{mE})L_{21}'(\alpha_{mE}),
\end{equation}
\begin{equation}
  B(\alpha_{mE})=L_{21}(\alpha_{mE})L_{12}'(\alpha_{mE})-L_{12}(\alpha_{mE})L_{21}'(\alpha_{mE}),
  \end{equation}
  and 
  \begin{equation}
  C(\alpha_{mE})=L_{22}(\alpha_{mE})L_{12}'(\alpha_{mE})-L_{12}(\alpha_{mE})L_{22}'(\alpha_{mE}),
  \end{equation}
respectively, where $L_{ij}'(\alpha)\equiv\partial L_{ij}/\partial \alpha$
denotes the derivative of coefficient $L_{21}$ evaluated at  $\alpha_{mP}$ and $\alpha_{mE}$ and the property $L_{22}=L_{11}$ was again used to derive Eq. (\ref{kame}).
The corresponding ${\cal P}_{mP,\alpha}$/$\eta_{mP,\alpha}$ is straightforwardly
evaluated and given by
\begin{equation}
   {\cal P}_{mP,\alpha}=\frac{TL_{12}'(\alpha_{mP})}{L_{11}'^2(\alpha_{mP})}[L_{12}(\alpha_{mP})L_{11}'(\alpha_{mP})-L_{11}(\alpha_{mP})L_{12}'(\alpha_{mP})]f_2^2,
\end{equation}
\begin{equation}
\eta_{mP,\alpha}=\frac{L_{12}'(\alpha_{mP})[L_{12}(\alpha_{mP})L_{11}'(\alpha_{mP})-L_{11}(\alpha_{mP})L_{12}'(\alpha_{mP})]}{L_{11}'(\alpha_{mP})[L_{22}(\alpha_{mP})L_{11}'(\alpha_{mP})-L_{21}(\alpha_{mP})L_{12}'(\alpha_{mP})]},
\end{equation}
respectively, and similar expressions are obtained for ${\cal P}_{mE,\alpha}$ and $\eta_{mE,\alpha}$
by inserting Eq. (\ref{kame}) into expressions
for ${\cal P}$ and $\eta$.
By focusing on the maximization with 
respect to the phase difference, we see that
the engine regime is delimited by two values of $\delta_{m1}$
and $\delta_{m2}$ in which ${\cal P}\ge 0$. From above
expressions,  the maxima $\delta_{mP}$ and $\delta_{mE}$ are given by
\begin{equation}
    \centering
    \delta_{mP}=\frac{1}{\omega}\tan ^{-1}\left\{\frac{-k^2-2 k \left(\kappa - \omega ^2\right)+ \omega ^2 \left[2 \kappa - \left(\omega ^2-\gamma ^2\right)\right]}{2 \gamma   \omega  \left(-\kappa
   -k+ \omega ^2\right)}\right\}
   \label{deltamP}
\end{equation}
and 
\begin{equation}
\frac{f_1}{f_2}=\frac{B(\delta_{mE})}{2L'_{21}(\delta_{mE})L_{11}}\left[1\mp \sqrt{1+\frac{4L^2_{11}L'_{21}(\delta_{mE})L'_{12}(\delta_{mE})}{B^2(\delta_{mE})}} \right],
\label{deltamE}
    \end{equation}
respectively.
We pause again to make some few comments: First, since the
lag appears only in crossed Onsager coefficients, the optimal
$\delta_{mP}$ does not depend on forces $f_1/f_2$,   solely depending on $\gamma,k,\kappa$ and $\omega$ [see e.g. dashed lines in Fig. \ref{fig5}$(b)$]. Second, for
$k+\kappa\gg \omega^2$ and $k+\kappa\ll \omega^2$, the optimal $\omega\delta_{mP} \rightarrow \pi/2$ and $-\pi/2$, respectively. Third, in contrast
with $\delta_{mP}$, $\delta_{mE}$ depends on ratio
$f_2/f_1$ [see e.g. dashed lines in Fig. \ref{fig5}$(a)$] and its value is given by the solution of transcendental
Eq. (\ref{deltamE}).
Fig. \ref{fig45} exemplifies the maximization of engine with respect
to the phase difference for some values of  output forces and Fig. \ref{fig5} shows (dashed lines), for several $f_1$ and $\delta$'s, 
the power and efficiency associate with the conversion
from  $\overline{\dot{W}}_2$ into  $\overline{\dot{W}}_1$ and vice-versa.

\subsubsection{Complete maximization of engine}
Here we address the optimization with
respect to the  output force and lag simultaneously. In other words,
 the maximum power output and efficiency must satisfy simultaneously Eqs. (\ref{x2mp})/(\ref{deltamP}) and  Eqs. (\ref{eq:x2meta})/(\ref{deltamE}), respectively. Starting with the
 power output, the existence of an optimal lag $\delta_{mP}^*$ and $f_{1mP}^*$ imply that
\begin{equation}
  \frac{L'_{12}(\delta_{mP}^*)}{L'_{11}(\delta_{mP}^*)} = \frac{1}{2}\frac{L_{12}(\delta_{mP}^*)}{L_{11}(\delta_{mP}^*)},
\end{equation}
and
\begin{equation}
  f_{1mP}^* = -\frac{1}{2}\frac{L_{12}(\delta_{mP}^*)}{L_{11}(\delta_{mP}^*)} f_2,
\end{equation}
respectively. Expressions for power and efficiency at maximum power at simultaneous maximizations are readily evaluated and given by
\begin{equation}
  {\cal P}^*_{mP} =  \frac{T}{4} \frac{L^2_{12}(\delta_{mP}^*)}{L_{11}(\delta_{mP}^*)} f_2^2, \label{eq:powboth}
\end{equation}
and
\begin{equation}
  \eta^*_{mP}=\frac{L^2_{12}(\delta_{mP}^*)}{4 L_{11}^2(\delta_{mP}^*)-2 L_{12}(\delta_{mP}^*)L_{21}(\delta_{mP}^*)}.
\end{equation}

Similar  expressions for the global  maximum efficiency and power at maximum efficiency are obtained by inserting  $f_{1mE}^*/\delta_{mE}^*$ into the expression for power 
and efficiency, respectively, the former being  given by
\begin{equation}
  \eta^*_{mE}=-\frac{L_{12}(\delta_{mE}^*)}{L_{21}(\delta_{mE}^*)}+\frac{2L_{11}^2(\delta_{mE}^*)}{L_{21}^2(\delta_{mE}^*)}\left(1-\sqrt{1-\frac{L_{21}(\delta_{mE}^*) L_{12}(\delta_{mE}^*)}{L_{11}^2 (\delta_{mE}^*)}}\right),
  \label{etame2}
  \end{equation}
  respectively.

Fig.  \ref{fig5} depicts the simultaneous
maximization of power and efficiency with respect
to the phase difference and output force for the same parameters from Fig; \ref{fig2}. For the sake of comparison, we also look a the lagless
case are depicted  in Fig. \ref{fig1}$(a)$ and $(b)$. 
Although the engine operates
rather inefficiently for $\delta=0$ (maximum efficiency and power read $\eta_{mE}\approx 0.172$ and 
${\cal P}_{mP}\approx 0.020$)
the simultaneous maximization of engine provides a substantial
increase of power and output, reading $\eta^*_{mE} \approx 0.382$ and ${\cal P}^*_{mP} \approx 0.081$. Similar findings are obtained for other values of $\kappa$
and $\omega$, in which the machine performance  increases
by raising $\kappa$ and lowering $\omega$.

\subsection{Different temperatures}\label{eng}
In this section, we derive general findings for the case of each particle   placed in contact with  a distinct thermal bath. We shall restrict
our analysis for $k+\kappa>\omega^2$, where
the efficiency is expected to be larger.
 Although the power output ${\cal P}$ 
is the same as before,  the efficiency may change due to the appearance of heat flow
and therefore its  maximization will occur
(in general) for distinct output forces and phase differences when
compared with the work-to-work converter.   The efficiency $\eta$
in such case  then reads:
   \begin{equation}
  \eta=-\frac{{\cal P}}{{\overline {\dot W_{\rm 2}}}+{\overline {\dot Q_{\rm i}}}}.
  \label{eff1}
\end{equation} 

Contrasting with the work-to-work converter, in which particles only dump heat to the reservoirs [and hence the heat is not  considered in Eq. \eqref{efff}], the temperature difference may be responsible for some amount of heat flowing from the reservoirs to
the system). As the power output is kept the same, the efficiency will always decrease as the temperature gap is raised.
For a small difference of temperatures, the heat regime occurs for a lower
range of  \(f_1\) or \(\delta\) than the entire engine regime, since $Q_{i}\le 0$ only for some specific
parameters. In other words,  let \(f_h\)  the threshold force separating both operation regimes (an analogous
description holds valid for $\delta_h$). For $|f_h|< |f_{1}|\le |f_m|$ the engine receives heat from one thermal bath, since \(\overline{\dot{Q}}_i<0\) or equivalently
${\overline \kappa}\Delta T -B_if_if_j> A_if_i^2+C_if_j^2$.
The force $f_h$ then satisfies \(\overline{\dot{Q}}_i (f_h) =0\), or equivalently $C_if_h^2+A_if_i^2={\overline \kappa}\Delta T-B_if_if_h$. For $0\le |f_{1}|\le |f_h|$, the machine then  works as a work-to-work converter and therefore the temperature difference is playing no role (results from Section~\ref{wtw} are held valid in this case).  It is worth mentioning that above inequality can be  satisfied under  distinct ways: for large $\Delta T$ and/or choices of $\delta$ or $f_1$. 


Despite all calculations being exact, expressions
for the efficiency and their maximizations become more involved, since they also depend on coefficients $A_i,B_i$
and $C_i$.
In order to obtain some insights about its behavior  in the presence of a heat flux, let us perform  an analysis for $\Delta T\ll1$
and $\Delta T\gg1$. In the former limit, $\eta$ is approximately given by
$\eta\approx -(\overline{\dot{W}}_1/\overline{\dot{W}}_2)(1-\overline{\dot{Q}}_i/\overline{\dot{W}}_2)$.
By expressing it in terms of Onsager coefficients,  one arrives at the following approximate expression for the efficiency
\begin{equation}
    \eta\approx-\frac{L_{11}f_1^2+L_{12}f_1f_2}{L_{22}f_2^2+L_{12}f_2f_1}\left(1+\frac{\overline{\dot{Q}}_i}{T(L_{22}f_2^2+L_{21}f_2f_1)}\right),
\end{equation}
where the input heat \(\overline{\dot{Q}}_i<0\) plays the role of decreasing the efficiency.
Maximizations with respect to $f_1$ and $\delta$ can be carried out from above (approximate) expression if $|f_{mE}|\ge |f_h|$ and $\delta_{mE}>\delta_h$ and from Eq. (\ref{eq:x2meta}) if $|f_{mE}| \le |f_h|$ and $\delta_{mE}\le \delta_h$.

For the opposite limit  $\Delta T\gg 1$, the
efficiency is approximately given by $\eta \approx -T(L_{12}f_1f_2+L_{11}f_1^2)/{\overline \kappa}\Delta T$, revealing that $\eta$ decreases asymptotically as \(\Delta T^{-1}\) for large temperature differences. Recalling that the numerator does not depend on the temperature (see e.g. Appendix \ref{appb}), it is clear that $\eta\ll 1$, with maximum values $\eta_{mE}$ and $\eta_{mE,\delta}$ given by $\eta_{mE}\approx{\cal P}_{mP}/{\overline \kappa}\Delta T$ and $\eta_{mE,\delta}\approx{\cal P}_{mP,\delta}/{\overline \kappa}\Delta T$ for $f_{1mP}$ and $\delta_{mP}$,
respectively. 
For an intermediate $\Delta T$, the system
receives heat from the hot thermal bath along $0<|f_1|<|f_m|$ or $\delta_{m1}\le \delta \le \delta_{m2}$, both   maximizations are straightforwardly calculated 
from Eq. (\ref{eff1}). 
Analogous relations are obtained for $T_i<T_j$ by replacing $\overline{\dot{Q}}_i$ for $\overline{\dot{Q}}_j$.
\begin{figure*}[ht]
    \centering
    \includegraphics[width=\textwidth]{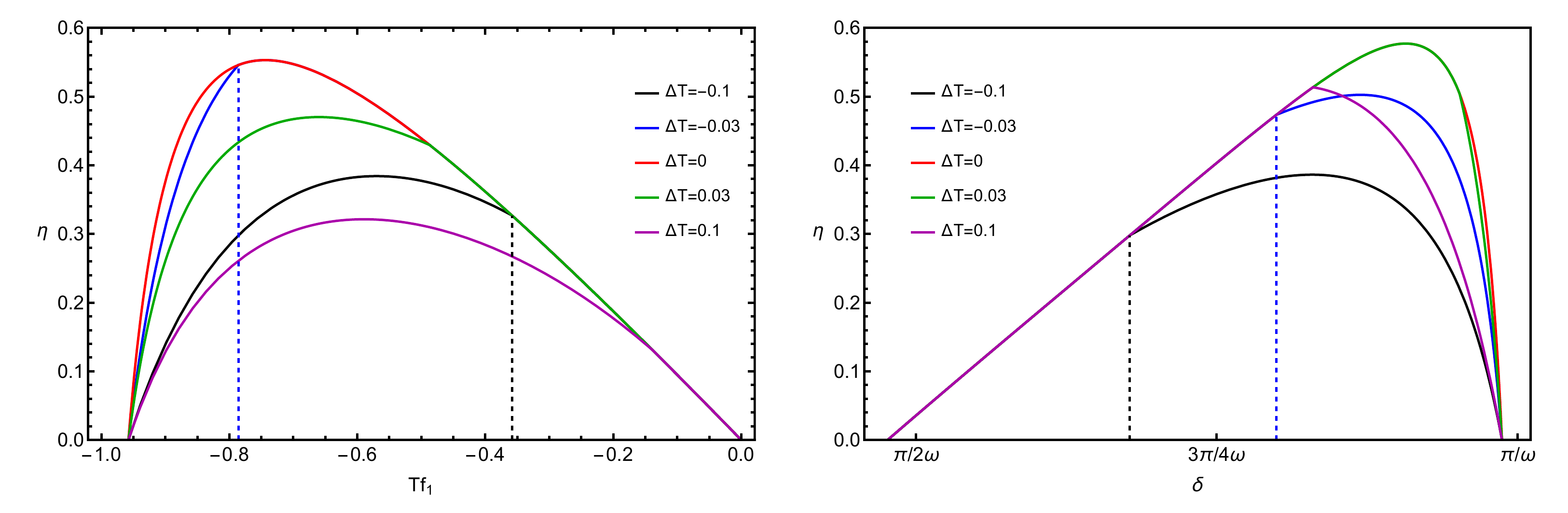}
    \caption{For distinct temperature reservoirs, left and right panels depict the efficiency versus \(Tf_1\) (for  \(\delta = 0\))  and versus \(\delta\) (for \(Tf_1 = 1\)), respectively.  The vertical lines denote the values of $f_h$  and $\delta_h$ separating the operation regimes. The red curves show the work-to-work efficiency. Parameters: $T=0.3+\Delta T/2,\omega=1$, $k=0.1$, $\kappa=5$ and $Tf_2=1$.}
    \label{eff_dT2}
\end{figure*}

\begin{figure*}[ht]
     \centering
     \includegraphics[scale=0.5]{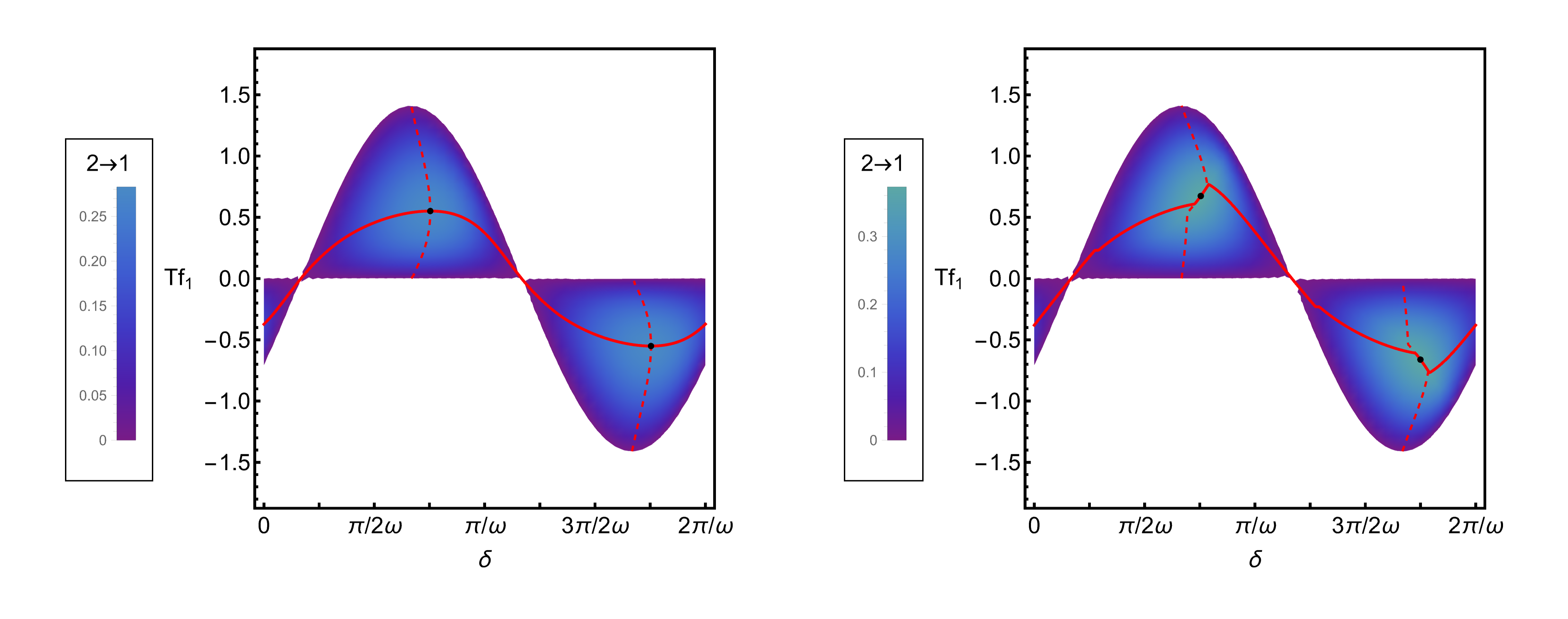}
     \caption{For the same parameters from Fig. \ref{fig5}, left and right panels depict of efficiency (for the conversion from $\overline{\dot{W}}_2$ into $\overline{\dot{W}}_1$) as a function of  $Tf_1$ and
      $\delta$ for $\Delta T=-0.3$ and $0.3$, respectively. Continuous and dashed lines denote
    the maximization with respect to the force $Tf_1$ and $\delta$, respectively. The simultaneous maximization (circles) corresponds to
    the intersection between maximum curves.}
     \label{fig9}
 \end{figure*}

In order
to illustrate above findings, Fig. \ref{eff_dT2} exemplifies the efficiency for distinct and small $\Delta T=T_2-T_1$ for fixed $\delta=0$ [left panel] and $f_1=1$ [right panel]. As
stated before, the power ${\cal P}$ is the same as in Fig. \ref{fig1}$(b)$ for $\kappa=5$.
 Since $\overline{\dot{Q}}_1$ and $\overline{\dot{Q}}_2$ exhibit 
distinct dependencies with $f_1$ and $\delta$, the amount of heat received will be different
when $\Delta T>0$ or $<0$. 
Such findings depict that it can more advantageous
 to receive
heat from the thermal bath 1 or 2 depending on the parameters the machine is projected. Such advantages are examined  in more details in Fig. \ref{fig9},
in which we extend for lower interaction parameter  and several values of $f_1$
and $\delta$ for $\Delta T=0.3$ and $-0.3$.
As for the work-to-work converter, there is also the global maximization corresponding to the intersection between both maximum lines.
Since the efficiency is lower than the work-to-work converter (see e.g. Fig. \ref{fig5}), the role
of the present optimization (whether with respect $f_1,\delta$ or both) 
reveals to be  relevant for enhancing the engine performance.


\section{Conclusions} \label{conc}
In this paper, we introduced and analyzed a model for a small scale engine based on interacting Brownian particles subject to periodically driving forces. General
expressions for the thermodynamic properties, power output and  efficiency 
were  investigated.
 Interaction between particles plays a central role not only for improving the machine performance but also for changing the machine regime operation. Furthermore, we observe the existence of distinct operation regimes for the same driving strength or phase difference. 
  The present framework reveals to be a  suitable route for obtaining efficient thermal engines that benefit from interactions and may constitute a first step for the description of  larger chains of interacting particles.
  It is worth pointing out  that positions and velocities get uncoupled for the sort of drivings we have considered and thereby the heat received by the particle  can not be converted into useful work.
   Hence, an interesting extension of the present work would be to exploit
  other kinds of time dependent drivings providing the
  heat to be  converted into useful work. 
  Another potential
  extension of our work would
  be to study engines composed of chains of larger systems sizes,  
  in order to compare the role of system size for enhancing the  efficiency and power.
\section{Acknowledgments}

Authors acknowledge financial support from S\~{a}o Paulo Research Foundation (FAPESP) under grants 2020/12021-6, 2017/24567-0, 2020/03708-8 and 2018/02405-1.


\onecolumngrid
\begin{center}
\textbf{\large Appendix}
\end{center}

\setcounter{equation}{0}
\setcounter{figure}{0}
\setcounter{table}{0}
\setcounter{page}{1}
\setcounter{section}{0}
\makeatletter
\renewcommand{\theequation}{A\arabic{equation}}
\renewcommand{\thefigure}{A\arabic{figure}}
\renewcommand{\citenumfont}[1]{A#1}

\subsection{Expressions for covariances}\label{appa}

 From the Fokker-Planck-Kramers equation, the time evolution of  covariances $b_{ij}^{xv}(t)\equiv \langle x_iv_j\rangle(t)-\langle x_i\rangle(t) \langle v_j\rangle(t)$ are given by
\begin{equation}
  \frac{d b_{11}^{xx}}{dt}  = 2b_{11}^{xv},
  \label{qw1}
\end{equation}
\begin{equation}
      \frac{d b_{11}^{vv}}{dt} = - 2 (k+\kappa)b_{11}^{xv} +2\kappa b_{12}^{xv} -2\gamma  b_{11}^{vv}+ 2\gamma  T_1,
\end{equation}
\begin{equation}
      \frac{d  b_{11}^{xv}}{dt}  =  b_{11}^{vv}- (k+\kappa)b_{11}^{xx}  + \kappa b_{12}^{xx} - \gamma b_{11}^{xv},
\end{equation}
\begin{equation}
      \frac{d  b_{12}^{xx}}{dt} =  b_{12}^{xv} + b_{21}^{xv},
\end{equation}
\begin{equation}
  \frac{d b_{12}^{vv}}{dt} = -  (k+\kappa) (b_{12}^{xv}+b_{21}^{xv}) +  \kappa (b_{11}^{xv}+b_{22}^{xv})   -  2 \gamma b_{12}^{vv},
  \label{qw2}
\end{equation}
and
\begin{equation}
    \frac{d b_{12}^{xv}}{dt}  =b_{12}^{vv}  - (k+\kappa) b_{12}^{xx} + \kappa b_{11}^{xx} - \gamma b_{12}^{xv},
\end{equation}
respectively, and analogous relations are obtained
for $b_{21}^{xx},b_{21}^{vv},b_{21}^{xv}$
and $b_{22}^{xx},b_{22}^{vv},b_{22}^{xv}$ just
by replacing $1\leftrightarrow 2$.  From the above set
of  linear equations, all  expressions for steady state covariances are obtained, as listed in Appendix \ref{appa}. Since only  $b_{ij}^{vv}$'s are needed for obtaining the entropy production, we shall omit their expressions, but they can be found in  Ref. \cite{tome2010entropy}.

\subsection{Expressions for the entropy production, average work and  heat 
 over a complete cycle}\label{appb}
In this appendix, we list the main expressions for  $\overline{\dot{W}}_1$, $\overline{\dot{W}}_2$ 
$\overline{\dot{Q}}_1$, $\overline{\dot{Q}}_2$ 
and $\overline{\sigma}$ averaged over a complete cycle. As stated previously,   our starting point are the relationships  $\dot{W}_i = -mF_i(t) \expval{v_i}$ and $\dot{Q}_i = \gamma\left( m\expval{v_i^2}-k_\text{B} T_i \right)$ together  averages $\langle v_i\rangle$'s and $\langle v_i\rangle^2$  integrated over a complete cycle.

The steady state entropy production  given by the expression 
\begin{equation}
  \overline{\sigma} =   \frac{\overline{\dot{Q}}_1}{T_1} +  \frac{\overline{\dot{Q}}_2}{T_2},
  \label{pi}
\end{equation}
which is a sum of  two terms: $\Phi_T$ and ${\overline \Phi_f}$. Such
latter one,  due to the external forces, has the
form $\tilde{L}_{11}X_1^2+(\tilde{L}_{12}+\tilde{L}_{21})X_1X_2+\tilde{L}_{22}X_2^2$, where coefficients (for $m=k_B = 1$) are given by
\begin{equation}
    \tilde{L}_{11} = \frac{\gamma \omega^2}{T_1T_2}\frac{T_1\kappa^2 + T_2 [(k+\kappa)^2 + \omega^2(\gamma^2 + \omega^2 -2(k+\kappa))]}{[\gamma^2\omega^2 + (\omega^2-k)^2 ] [\gamma^2\omega^2 + (\omega^2-k-2\kappa)^2 ]},
\end{equation}
\begin{equation}
    \tilde{L}_{12}+\tilde{L}_{21} = \frac{\gamma \omega^2 \kappa}{2T_1T_2}\frac{ (T_1+T_2)(k+\kappa-\omega^2)\cos(\delta \omega) + (T_1-T_2)\gamma \omega \sin(\delta\omega)}{[\gamma^2\omega^2 + (\omega^2-k)^2 ] [\gamma^2\omega^2 + (\omega^2-k-2\kappa)^2 ]},
\end{equation}
and
\begin{equation}
    \tilde{L}_{22} = \frac{\gamma \omega^2}{T_1T_2}\frac{T_2\kappa^2 + T_1 [(k+\kappa)^2 + \omega^2(\gamma^2 + \omega^2 -2(k+\kappa))]}{[\gamma^2\omega^2 + (\omega^2-k)^2 ] [\gamma^2\omega^2 + (\omega^2-k-2\kappa)^2 ]},
\end{equation}
respectively. Note that above coefficients reduce
to Onsager coefficients $L_{21}$'s when $T_1=T_2$.

In order to relate coefficients ${\tilde L}_{ij}$'s  with Onsager ones   $L_{ij}$'s, it is convenient to expand Eq. (\ref{eq:phi}) 
in the regime of small $\Delta T$, in such a way that $ \overline{\sigma}$ is approximately given by
\begin{equation}
   \overline{\sigma}\approx  \left[ -\frac{1}{T}\left(\overline{\dot{W}}_1 + \overline{\dot{W}}_2\right) +  \left (\overline{\dot{Q}}_1 - \overline{\dot{Q}}_2\right) \frac{\Delta T}{2T^2}\right].
   \label{eq:phi2}
\end{equation}
Since the dependence with $\Delta T$
is present only in the second right term, it is clear that  Onsager coefficients $L_{21}$'s ($i,j\in 1,2$) correspond to 0-th order
    coefficients obtained from the expansion of $\overline{\sigma}$.
For this reason, the coefficient ${\tilde L}_{ij}$ can be decomposed as ${\tilde L}_{ij} = L_{ij} + L_{ij}^{(c)} \Delta T$,
where  $L_{ij}^{(c)}$ is the first order correction and then $\overline{\sigma}$ is given by
\begin{equation}
  \begin{split}
    \label{eq:pi}
   \overline{\sigma} 
   &\approx   L_{11} f_1^2 + \left(L_{12}+L_{21} \right) f_1 f_2 + L_{22} f_2^2 +\\
   & \left[ L_{11}^{(c)} f_1^2 + \left(L_{12}^{(c)}+L_{21}^{(c)} \right) f_1 f_2 + L_{22}^{(c)} f_2^2 \right] \Delta T+L_{TT}f_T^2,
  \end{split}
\end{equation}
where  $L_{TT}=\overline{\kappa}T^2>0$  with  $f_1=X_1/T$, $f_2=X_2/T$ and $f_T=\Delta T/T^2$  [where $T=(T_1+T_2)/2$].
As analyzed in Sec. II, for small $\Delta T$ and  $f_i$'s, the difference between $L_{ij}$'s and ${\tilde L}_{ij}$'s  can be neglected and the entropy production is approximately given by $\overline{\sigma}\approx   L_{11} f_1^2 + \left(L_{12}+L_{21} \right) f_1 f_2 + L_{22} f_2^2+L_{TT}f_T^2$.

The averaged expressions for $\overline{\dot{W}}_1$, $\overline{\dot{W}}_2$, $\overline{\dot{Q}}_1$
and $\overline{\dot{Q}}_2$ are given by
\begin{align}
\label{work}
    \overline{\dot{W}}_1&=-\frac{ T^2\gamma   \omega ^2\left( \gamma^2 \omega^2 + \left(\omega^2 -(k+ \kappa)  \right)^2 + \kappa^2\right) }{2\left[\gamma^2\omega^2 +\left(\omega^2 - k\right)^2  \right]  \left[ \gamma^2\omega^2 + (\omega^2 - (k+2\kappa))^2 \right]}f_1^2\nonumber\\
   &-\frac{T^2\kappa   \omega}{2}\frac{2 \gamma  \omega  \left(\kappa +k- \omega ^2\right)\cos (\delta  \omega )-  \sin (\delta  \omega ) \left[\gamma^2\omega^2 - (\omega^2 - (k+\kappa))^2 + \kappa^2  \right]}{\left[\gamma^2\omega^2 +\left(\omega^2 - k\right)^2  \right]  \left[ \gamma^2\omega^2 + (\omega^2 - (k+2\kappa))^2 \right]}f_1f_2,
   \end{align}
   \begin{align}
    \overline{\dot{W}}_2&=-\frac{T^2\gamma   \omega ^2\left(  \gamma^2 \omega^2 + \left(\omega^2 -(k+ \kappa)  \right)^2 + \kappa^2\right) }{2\left[\gamma^2\omega^2 +\left(\omega^2 - k\right)^2  \right]  \left[ \gamma^2\omega^2 + (\omega^2 - (k+2\kappa))^2 \right]}f_2^2\nonumber\\
   &-\frac{T^2\kappa   \omega}{2}\frac{2 \gamma  \omega  \left(\kappa +k- \omega ^2\right)\cos (\delta  \omega )+  \sin (\delta  \omega ) \left[\gamma^2\omega^2 - (\omega^2 - (k+\kappa))^2 + \kappa^2  \right]}{\left[\gamma^2\omega^2 +\left(\omega^2 - k\right)^2  \right]  \left[ \gamma^2\omega^2 + (\omega^2 - (k+2\kappa))^2 \right]}f_1f_2,
   \end{align}
  where
 \begin{align}
     \centering
    \overline{\dot{Q}}_1&=\frac{T^2\gamma  \omega ^2 \left[\gamma ^2 \omega ^2+\left(\kappa +k-\omega ^2\right)^2\right]}{2 \left[\gamma ^2 \omega ^2+\left(k-\omega ^2\right)^2\right] \left[\gamma ^2 \omega ^2+\left(2 \kappa +k-\omega ^2\right)^2\right]}f_1^2+\frac{T^2\gamma  \kappa ^2 \omega ^2}{2 \left[\gamma ^2 \omega ^2+\left(k-\omega ^2\right)^2\right] \left[\gamma ^2 \omega ^2+\left(2 \kappa +k-\omega ^2\right)^2\right]}f_2^2\nonumber\nonumber\\
    &+\frac{T^2\gamma  \kappa  \omega ^2 \left[\cos (\delta  \omega ) \left(\kappa +k-\omega ^2\right)-\gamma  \omega  \sin (\delta  \omega )\right]}{\left[\gamma ^2 \omega ^2+\left(k-\omega ^2\right)^2\right] \left[\gamma ^2 \omega ^2+\left(2 \kappa +k-\omega
   ^2\right)^2\right]}f_1f_2+\frac{\gamma \kappa ^2 }{2[\gamma ^2 k +\kappa
    \left(\kappa +\gamma ^2 \right)]} \Delta T
    \label{q1}
    \end{align}
    and
    \begin{align}
    \centering
   \overline{\dot{Q}}_2&=\frac{T^2\gamma  \kappa ^2 \omega ^2}{2 \left[\gamma ^2 \omega ^2+\left(k-\omega ^2\right)^2\right] \left[\gamma ^2 \omega ^2+\left(2 \kappa +k-\omega ^2\right)^2\right]}f_1^2+\frac{T^2\gamma  \omega ^2 \left[\gamma ^2 \omega ^2+\left(\kappa +k-\omega ^2\right)^2\right]}{2 \left[\gamma ^2 \omega ^2+\left(k-\omega ^2\right)^2\right] \left[\gamma ^2 \omega ^2+\left(2 \kappa +k-\omega ^2\right)^2\right]}f_2^2\nonumber\\
    &+\frac{T^2\gamma  \kappa  \omega ^2 \left[\cos (\delta  \omega ) \left(\kappa +k-\omega ^2\right)+\gamma  \omega  \sin (\delta  \omega )\right]}{\left[\gamma ^2 \omega ^2+\left(k-\omega ^2\right)^2\right] \left[\gamma ^2 \omega ^2+\left(2 \kappa +k-\omega
   ^2\right)^2\right]}f_1f_2-\frac{\gamma  \kappa ^2 }{2\left[\gamma ^2 k +\kappa
    \left(\kappa +\gamma ^2 \right)\right]} \Delta T
    \label{q2}
\end{align}
respectively.
\twocolumngrid
\bibliography{refs}
\end{document}